\documentclass[prd,
reprint, 
groupedaddress,
showpacs,
nofootinbib,
 amsmath,amssymb,
 aps,
floats,
floatfix,
twocolumn]{revtex4-1}

\usepackage{graphicx}
\usepackage{dcolumn}
\usepackage{bm}
\usepackage{url}
\newcommand{\PR}[1]{\ensuremath{\left[#1\right]}} 
\newcommand{\PC}[1]{\ensuremath{\left(#1\right)}} 
\usepackage{hyperref}

\begin{document}


\title{Constraints on hybrid metric-Palatini models from background evolution}

\author{Nelson A. Lima}
 \email{ndal@roe.ac.uk}
\author{Vanessa Smer-Barreto}
 \email{vsm@roe.ac.uk}
 \affiliation{Institute for Astronomy, University of Edinburgh, Royal Observatory, Blackford Hill, Edinburgh, EH9 3HJ, UK}

\renewcommand{\abstractname}{Abstract}
\begin{abstract}
In this work, we introduce two models of the hybrid metric-Palatini theory of gravitation. We explore their background evolution, showing explicitly that one recovers standard General Relativity with an effective Cosmological Constant at late times. This happens because the Palatini Ricci scalar evolves towards and asymptotically settles at the minimum of its effective potential during cosmological evolution. We then use a combination of cosmic microwave background, supernovae and baryonic accoustic oscillations background data to constrain the models' free parameters. For both models, we are able to constrain the maximum deviation from the gravitational constant $G$ one can have at early times to be around $1\%$.
\end{abstract}

\pacs{98.80.-k, 95.36.+x, 04.50.Kd \hfill \today}
\maketitle


\section{\label{Int}Introduction}

The $\Lambda$CDM model, which provides the simplest explanation for the Universe's accelerated expansion in the form of a cosmological constant $\Lambda$ (for a review on this topic, see \cite{lambdareview}), is the main framework of modern cosmology. Despite being in agreement with supernovae observations \cite{accel1,accel2,accel3,accel4}, data from the cosmic microwave background (CMB) \cite{cmb1,cmb2} including the recent {\it Planck} data \cite{planck2015}, and large-scale structure (LSS) data \citep{lss1}, cosmologists still struggle to account for the difference between the theoretically-expected value for its energy density and the observed one.

In light of these issues, a new physical model may be in order to account for that major component of our Universe, usually labeled dark energy (DE). Theories that obey the standard gravitational field equations of General Relativity (GR), such as quintessence, k-essence, and so on, propose scalar fields rolling in a potential (see \cite{quintessence} and references therein for a comprehensive review). However, a much larger set of models arise from investigations where GR is assumed to fail, therefore proposing corrections to Einstein's action. These are grouped under the name of Modified Gravity Theories (MGT), such as the Brans--Dicke scalar--tensor theory \cite{bd}, higher dimensions like in braneworld models \cite{DGP, branes}, Galileon models \cite{gall}, the Fab Four \cite{fabfour}, $f(R)$ theories \cite{frreview},  and many others. For an extensive review on MGT, see \cite{reviewall}. 

Of particular relevance to the background of the present work are $f(R)$ theories. In these models, GR is modified by taking the Lagrangian density to be a nonlinear function $f$ of the Ricci scalar. The dynamics of this metric variational approach can be interpreted as due to a Brans-Dicke scalar field with parameter $w=0$ and a non-trivial potential $V(\phi)$. However, when Solar System constraints are imposed, the interaction of this scalar field is restrained to a milimetric scale, nullifying its impact on cosmology \cite{quantization}. 

Another class of gravity theories, known as Palatini, emanate from considering the metric and the connection as independent variables \cite{DEbook}. However, these models have been observed to be unstable to matter loop corrections. If the matter loop effects are accounted for, they give rise to a Lagrangian density for the gravitational action which is a function of the Ricci scalars of both the connection and the metric, which in turn creates a set of theories named hybrid-metric Palatini \cite{main1,main2}. The latter are the focus of this paper, as we deem interesting to explore the late time acceleration effects of a set of models who consider perturbative quantization methods \cite{quantization} whilst also displaying connections with non-perturbative quantum geometries \cite{main1}. 

Like the pure metric and Palatini cases, the hybrid theory has a dynamically equivalent scalar-tensor representation \cite{main1,main2}. The authors of those papers have also shown that the scalar field need not be massive in order to pass the stringent Solar System constraints \cite{main1}, in contrast to the metric $f(R)$ theories, while possibly modifying the cosmological \cite{hybridcosmo} and Galactic \cite{hybridgala} dynamics due to its light, long-range interacting nature. It has actually been shown that the additional scalar degree of freedom introduced by this theory does not need to become massive in order to evade solar-system tests. The viability of the theory on the smallest scales can be assured as long as the background value of the scalar field remains small \cite{main1,main2}. Hence, this theory seems to not need an explicit screening mechanism, even though much work remains to be done on this topic.

The criteria for obtaining cosmic acceleration have already been discussed and introduced \cite{hybridcosmo}. Alongside that, several cosmological solutions were derived, depending on the form of the effective scalar field potential, describing both accelerating and decelerating Universes. Lastly, the full set of linearly perturbed Einstein equations was derived for this theory, and the evolution of the metric potentials was shown for a designer model reproducing the $\Lambda$CDM background evolution exactly \cite{hybridpert}.

In this work, we show the background history predicted by two $f(\mathcal{R})$ models we introduce, which is achieved by numerically evolving the Palatini Ricci scalar, $\mathcal{R}$, that is then used to compute the remaining background quantities. We then test the models against background data, using this to constrain their free parameters. Therefore, in Section \ref{II}, we briefly review the hybrid metric-Palatini theory formalism, focusing on the background equations we will be using throughout this work. In Sec.~\ref{III}, we introduce and dissect the two models we have used in this study, showing the background evolution predicted by these against $\Lambda$CDM. In Section \ref{IV} we present the background constraints on our models using a MCMC analysis, and finish in Sec.~\ref{conclusion} with some concluding remarks on this work.

\section{\label{II}Cosmology in the hybrid metric-Palatini gravity}

The four-dimensional action describing the hybrid metric-Palatini gravity is given by
\begin{equation}{\label{action}}
 S = \frac{1}{2\kappa^2} \int d^4 x \sqrt{-g}\PR{R + f(\mathcal{R})} + \int d^4x \sqrt{-g} \mathcal{L}_{\rm{m}},
\end{equation}
\noindent where $\kappa^2 = 8 \pi G$ and we set $c=1$. $\mathcal{L}_{\rm{m}} = \mathcal{L}_{\rm{m}}(g^{\mu \nu},\psi)$ is the standard minimally--coupled matter Lagrangian, $R$ is the metric Einstein-Hilbert Ricci scalar and $\mathcal{R} = g^{\mu \nu} \mathcal{R}_{\mu \nu}$ is the Palatini curvature. The latter is defined in terms of the metric elements, $g^{\mu \nu}$, and a torsion-less independent connection, $\hat{\Gamma}$, through
\begin{equation}
\mathcal{R} \equiv g^{\mu \nu}\PC{\hat{\Gamma}^{\alpha}_{\mu \nu, \alpha} - \hat{\Gamma}^{\alpha}_{\mu \alpha,\nu} + \hat{\Gamma}^{\alpha}_{\alpha \lambda} \hat{\Gamma}^{\lambda}_{\mu \nu} - \hat{\Gamma}^{\alpha}_{\mu \lambda} \hat{\Gamma}^{\lambda}_{\alpha \nu}}.
\end{equation}

Varying the action (\ref{action}) with respect to the metric, one obtains the usual set of Einstein equations, given by
\begin{equation}{\label{einstein}}
 G_{\mu \nu} + F(\mathcal{R})\mathcal{R}_{\mu \nu} - \frac{1}{2}f(\mathcal{R})g_{\mu \nu} = \kappa^2 T_{\mu \nu},
\end{equation}
\noindent where $G_{\mu \nu}$ is Einstein's tensor, and $F(\mathcal{R}) \equiv df(\mathcal{R})/d\mathcal{R}$. $T_{\mu \nu}$ is the matter field's stress--energy tensor, defined as
\begin{equation}
 T_{\mu \nu} = -\frac{2}{\sqrt{-g}} \frac{\delta\PC{\sqrt{-g} \mathcal{L}_{\rm{m}}}}{\delta \PC{g^{\mu \nu}}}.
\end{equation}
\noindent Equation~(\ref{einstein}) can be traced, yielding
\begin{equation}{\label{einsteintrace}}
 R + \kappa^{2} T= F(\mathcal{R})\mathcal{R} - 2f(\mathcal{R}),
\end{equation}
\noindent where $T$ is the stress--energy tensor trace. The last equation shows that the modifications to Einstein gravity are controlled by the failure of the standard GR trace equation.

On the other hand, varying the action (\ref{action}) with respect to the independent connection, $\hat{\Gamma}^{\alpha}_{\mu \nu}$, one gets the following equation,
\begin{equation}{\label{independent}}
 \hat{\nabla}_{\alpha}\PC{\sqrt{-g}F(\mathcal{R})g^{\mu \nu}} = 0,
\end{equation}
\noindent which implies that $\hat{\Gamma}^{\alpha}_{\mu \nu}$ is the Levi--Civita connection of a metric $h_{\mu \nu} = F(\mathcal{R})g_{\mu \nu}$. Therefore, $h_{\mu \nu}$ is a conformal transformation of the original metric elements $g_{\mu \nu}$ by the factor $F(\mathcal{R})$.  Hence, one can verify that the Palatini Ricci tensor, $\mathcal{R}_{\mu \nu}$, is related to the metric one, $R_{\mu \nu}$, by the following relation
\begin{eqnarray}{\label{riccirelation}}
  \mathcal{R}_{\mu \nu} &=& R_{\mu \nu} + \frac{3}{2} \frac{1}{F^{2}(\mathcal{R})}F(\mathcal{R})_{,\mu}F(\mathcal{R})_{,\nu} - \nonumber \\
  &-& \frac{1}{F(\mathcal{R})}\nabla_{\mu}F(\mathcal{R})_{,\nu} - \frac{1}{2}\frac{1}{F(\mathcal{R})}g_{\mu \nu} \Box F(\mathcal{R}).
\end{eqnarray}
\noindent Adopting a flat Friedmann-Robertson-Walker (FRW) metric, $ds^2 = - dt^2 + a^{2}(t)d\vec{x}^2$, one can get the modified Friedmann equations:
\begin{equation}{\label{hubble}}
 3H^{2} = \frac{1}{1 + f_{\mathcal{R}}} \PR{\kappa^{2} \rho - 3 H \dot{f_{\mathcal{R}}} - \frac{3 \dot{f_{\mathcal{R}}}^{2}}{4 f_{\mathcal{R}}} + \frac{\mathcal{R} F(\mathcal{R}) - f(\mathcal{R})}{2}},
\end{equation}
\noindent and,
\begin{equation}{\label{hubbledot}}
 2\dot{H} = \frac{1}{1 + f_{\mathcal{R}}} \PR{ -\kappa^{2}\PC{\rho + p} + H \dot{f_{\mathcal{R}}} - \ddot{f_{\mathcal{R}}} + \frac{3 \dot{f_{\mathcal{R}}}^{2}}{2 f_{\mathcal{R}}}},
\end{equation}
\noindent where a dot stands for a differentiation with respect to time, $t$, and $H = \dot{a}/a$ is the Hubble parameter; $\rho$ and $p$ are the total energy density and pressure of the Universe. An equivalent equation for $\dot{H}$ can be taken from the trace of Eq.~(\ref{einstein}), yielding
\begin{equation}{\label{hubbledotmain}}
 \dot{H} = \frac{1}{6}\PR{F(\mathcal{R})\mathcal{R} - 2 f(\mathcal{R}) - \kappa^2 T - 12 H^{2}},
\end{equation}
\noindent where we have used $R = 6\PC{\dot{H} + 2H^{2}}$.



A scalar-tensor description of this theory has been developed, where the additional scalar degree of freedom is identified as $\phi = F(\mathcal{R})$ \cite{main1,main2}. Eq. (11) in \cite{main1} is an effective Klein-Gordon equation for the scalar field. Using the FRW metric, it takes the form

\begin{equation}{\label{dynamicalscalar}}
 \ddot{\phi} + 3H\dot{\phi} - \frac{\dot{\phi}^{2}}{2\phi} + \frac{\phi}{3}\PR{2V - \PC{1+\phi}\frac{dV}{d\phi}} = \frac{\kappa^2 \phi}{3}T,
\end{equation}
\noindent Equation (\ref{dynamicalscalar}) can be re-expressed as a dynamical equation for the Palatini Ricci scalar, $\mathcal{R}$. Recalling that $\phi \equiv F(\mathcal{R}) \equiv f_{\mathcal{R}}$, then one can set $\dot{F} = \dot{\mathcal{R}} f_{\mathcal{R} \mathcal{R}}$, where $f_{\mathcal{R} \mathcal{R}}$ is the second derivative of $f(\mathcal{R})$ with respect to $\mathcal{R}$. A similar procedure can be done for higher order derivatives, allowing to rewrite  Eq.~(\ref{dynamicalscalar}) as
\begin{eqnarray}{\label{dynamicalpala}}
  \ddot{\mathcal{R}} &=& - \frac{1}{f_{\mathcal{R}\mathcal{R}}} \Bigg[ \dot{\mathcal{R}}^{2}\PC{f_{\mathcal{R}\mathcal{R}\mathcal{R}} - \frac{f^{2}_{\mathcal{R}\mathcal{R}}}{2 f_{\mathcal{R}}}} + 3 H \dot{\mathcal{R}} f_{\mathcal{R}\mathcal{R}} + \nonumber \\
  &+& \frac{f_{\mathcal{R}}}{3}\PR{\mathcal{R}\PC{f_{\mathcal{R}}-1} - 2 f(\mathcal{R})} - \kappa^{2} \frac{f_{\mathcal{R}}}{3} T \Bigg]
\end{eqnarray}
\noindent where we have used Eq.~(5) in \cite{main1}, and $f_{\mathcal{R}\mathcal{R}\mathcal{R}}$ is the third order derivative of $f(\mathcal{R})$ with respect to $\mathcal{R}$. From this equation, we can define an effective potential where the effective Palatini Ricci scalar will roll, since
\begin{equation}{\label{effective_V_r}}
 \frac{dV_{\rm{eff}}}{d \mathcal{R}} = \frac{f_{\mathcal{R}}}{3f_{\mathcal{R}\mathcal{R}}}\PR{\mathcal{R}\PC{f_{\mathcal{R}}-1} - 2 f(\mathcal{R})},
\end{equation}
\noindent meaning $V(\mathcal{R})$ will be given by the indefinite integral
\begin{equation}{\label{effective_pot}}
 V(\mathcal{R}) = \int^{\mathcal{R}} \frac{f_{\mathcal{R}}}{3f_{\mathcal{R}\mathcal{R}}}\PR{\mathcal{R}\PC{f_{\mathcal{R}}-1} - 2 f(\mathcal{R})} d\mathcal{R}.
\end{equation}
It was shown that the hybrid metric-Palatini theory reduces to General Relativity with a possible cosmological constant in vacuum, since it shares the property of pure Palatini $f(R)$ theories in Minkowski flat space-time \cite{flatstable}. Furthermore, the field's equation of motion have been analyzed as a dynamical system: it was explicitly shown that as long as one provides a suitable $V(\phi)$, or equivalently a function $f(\mathcal{R})$ such that the slope of the potential is downwards and its minimum happens for a small value of the scalar field $\phi$, one should always obtain a natural transition from standard cosmological evolution to accelerated expansion close to the present while also avoiding any conflict with solar-system constraints \cite{hybridcosmo}. 

Hence, we use Eqs.~(\ref{hubbledotmain}) and (\ref{dynamicalpala}) to numerically evolve the background quantities predicted by a specific $f(\mathcal{R})$ model. To set the initial conditions, we have fixed a very small value for $f_{\mathcal{R}}$ at a high redshift, $z_{\rm{i}}$, such that the deviation from the Gravitational constant, $G$, is effectively small \cite{main1,main2,hybridgala,hybridpert} in the high curvature regime. Then, one can invert $F$ to find $\mathcal{R}$ at that redshift.

Lastly, we set $\dot{\mathcal{R}_{\rm{i}}} = 0$ to minimize deviations from standard General Relativity and use Eq.~(\ref{hubble}) to solve for the Hubble parameter at $z_{\rm{i}}$. The Ricci scalar, $R$, can be computed using the GR relation, $R = 6 \PC{\dot{H} + 2 H^{2}}$. Even though $\dot{\mathcal{R}_{\rm{i}}} = 0$ is a strong assumption, we have tested the models for a fairly broad range of initial velocities, within a slow-roll regime, and observed that their qualitative behavior remained unaltered with the late-time evolution tending to an effective Cosmological Constant, and $\mathcal{R}$ asymptotically reaching the equilibrium position predicted by the effective potential we define in Eq. (\ref{effective_pot}).

\section{\label{III}Models of hybrid metric-Palatini gravity}

In this section, we introduce the models we defined for the hybrid metric-Palatini theory. While the general framework of this theory was derived in Refs.~\cite{main1,main2,hybridcosmo,hybridgala}, they did not write down specific models and explore their consequences. These were inspired in theories of $f(R)$ gravity, such as the Starobinsky \cite{Starob} and the exponential \cite{expo} models, but are essentially phenomenological choices that are simply tested with the background evolution formalism displayed before.

\subsection{\label{expo}The exponential model}

The first model we introduce is defined by an exponential function, given by
\begin{equation}{\label{exponentialf}}
 f(\mathcal{R}) = \Lambda_{\star}\PC{1 + e^{-\mathcal{R}/\mathcal{R}_{\star}}},
\end{equation}
\noindent where $\Lambda_{\star}$ and $\mathcal{R}_{\star}$ are the model's parameters, both of order $H_{0}^{2}$, where $H_{0}$ is the present-day value of the Hubble parameter. We choose to define $\mathcal{R}_{\star}$ as a positive constant, while $\Lambda_{\star}$ should be a negative constant, since it should dominate over the other corrective terms introduced by our models. This is particularly relevant at late-times where, to recover standard GR with an effective Cosmological Constant, one should have $\Lambda_{\star} \approx -2 \Lambda$. Hence, $\Lambda_{\star} < 0$ allows the effective Cosmological Constant to have the correct sign, which becomes clear, for instance, looking at Eq. (\ref{einsteintrace}) in vacuum.

The effective potential, $V(\mathcal{R})$, associated to this model, is given by doing the indefinite integral defined by Eq.~(\ref{effective_pot}), which has the simple form
\begin{eqnarray}{\label{expopot}}
  V(\mathcal{R}) &=& \frac{\mathcal{R}_{\star}}{3} \Bigg[ 2\Lambda_{\star}\mathcal{R} + \frac{1}{2}\mathcal{R}^{2} - 2 \Lambda_{\star} \mathcal{R}_{\star} e^{-\mathcal{R}/\mathcal{R}_{\star}} -\nonumber\\
  &-&  e^{-\mathcal{R}/\mathcal{R}_{\star}}\PC{\Lambda_{\star}\mathcal{R} + \Lambda_{\star} \mathcal{R}_{\star}} \Bigg].
\end{eqnarray}
 We can see that the potential will be mostly dominated by its quadratic terms for order unit values of $\mathcal{R}_{\star}$. The minimum can then be estimated by
\begin{equation}{\label{expomin}}
 \mathcal{R}_{\rm{min}} \approx -2\Lambda_{\star}.
\end{equation}
\noindent The effective cosmological constant value at which the Ricci scalar should sit in vacuum is given by the trace equation
\begin{equation}{\label{expoefflamda}}
 \Lambda_{\rm{eff}} = \frac{1}{2} \PR{-2\Lambda_{\star}f_{\mathcal{R}}(-2\Lambda_{\star})-2f(\Lambda_{\star})} \approx -\Lambda_{\star}.
\end{equation}
\noindent Hence, we do expect to obtain a $\Lambda$CDM like evolution in the distant future, if our solution for $\mathcal{R}$ is to settle at the minimum of its potential.

We choose a small value for $f_{\mathcal{R}}$ (less than unit), independently of the starting redshift $z_{\rm{i}}$. The reason for this is, as mentioned in the end of Sec.~\ref{II}, that $f_{\mathcal{R}}$ sets the initial deviation from the actual gravitational constant in the high curvature regime. Therefore, higher values of $f_{\mathcal{R}}$ will either suppress gravity considerably or invert its sign if it is smaller than $-1$.
 
\begin{figure}[t!]
\begin{center}$
\begin{array}{c}
\includegraphics[scale = 0.31]{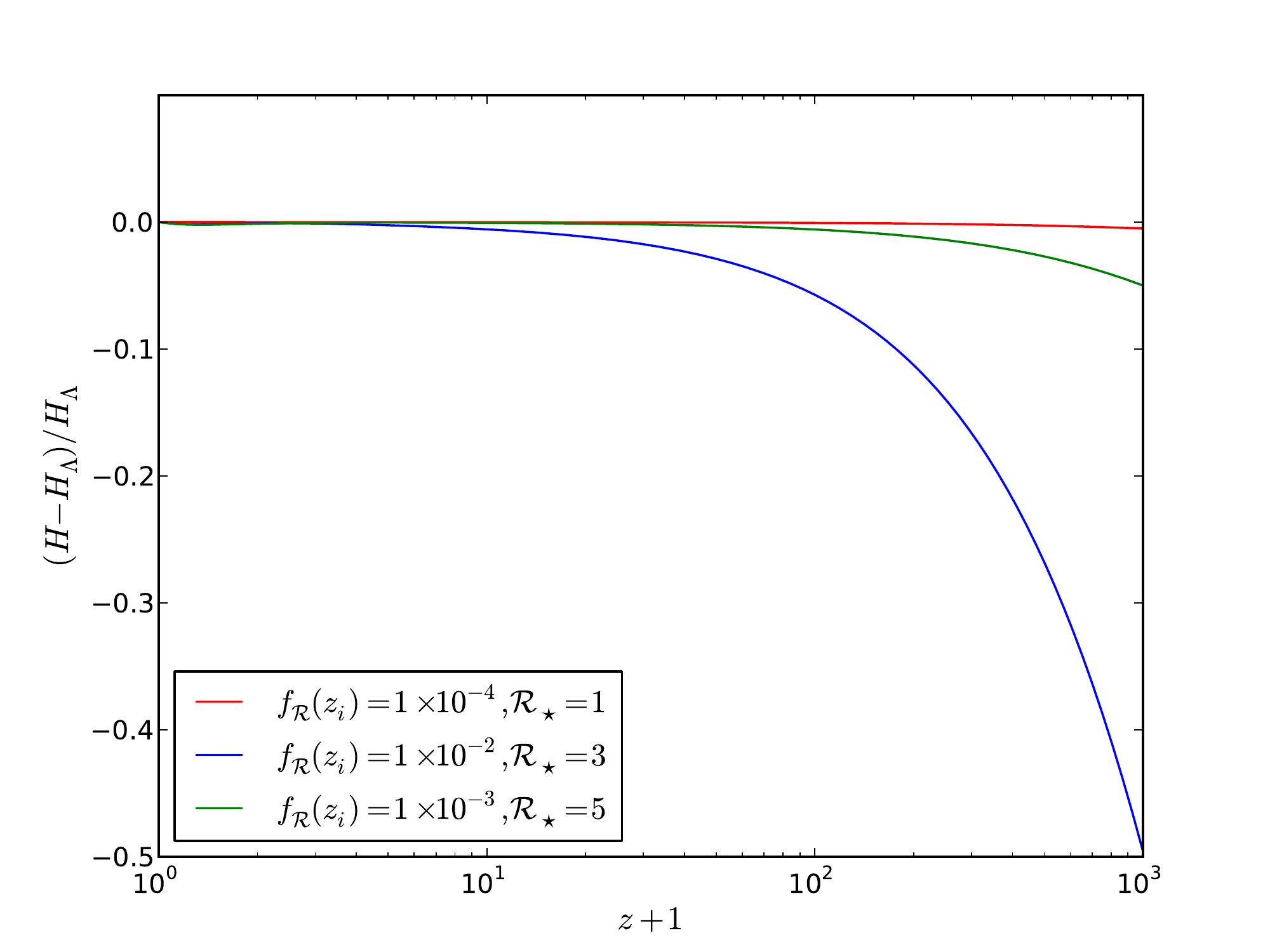} \\
\includegraphics[scale = 0.31]{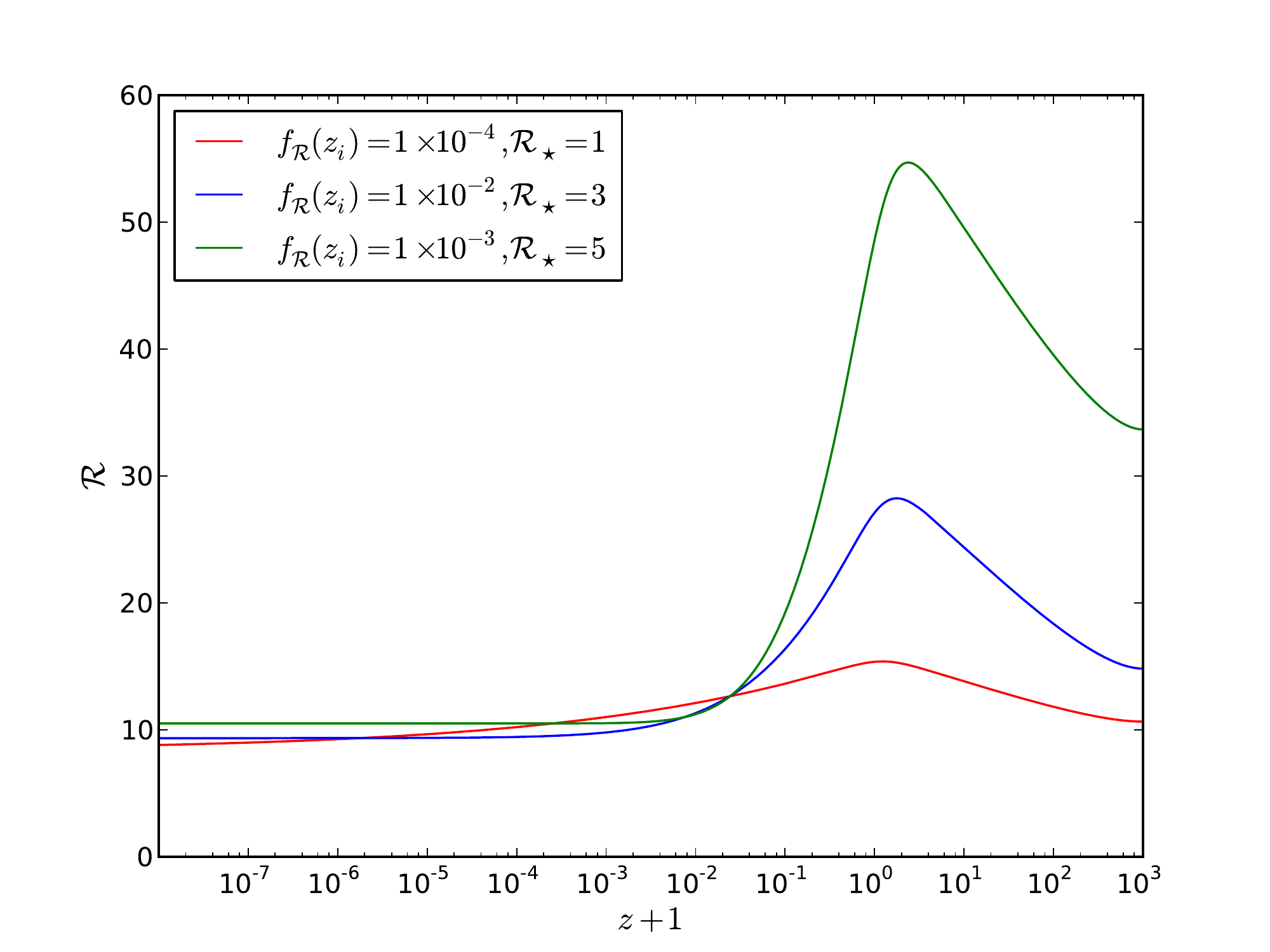} \\
\includegraphics[scale = 0.31]{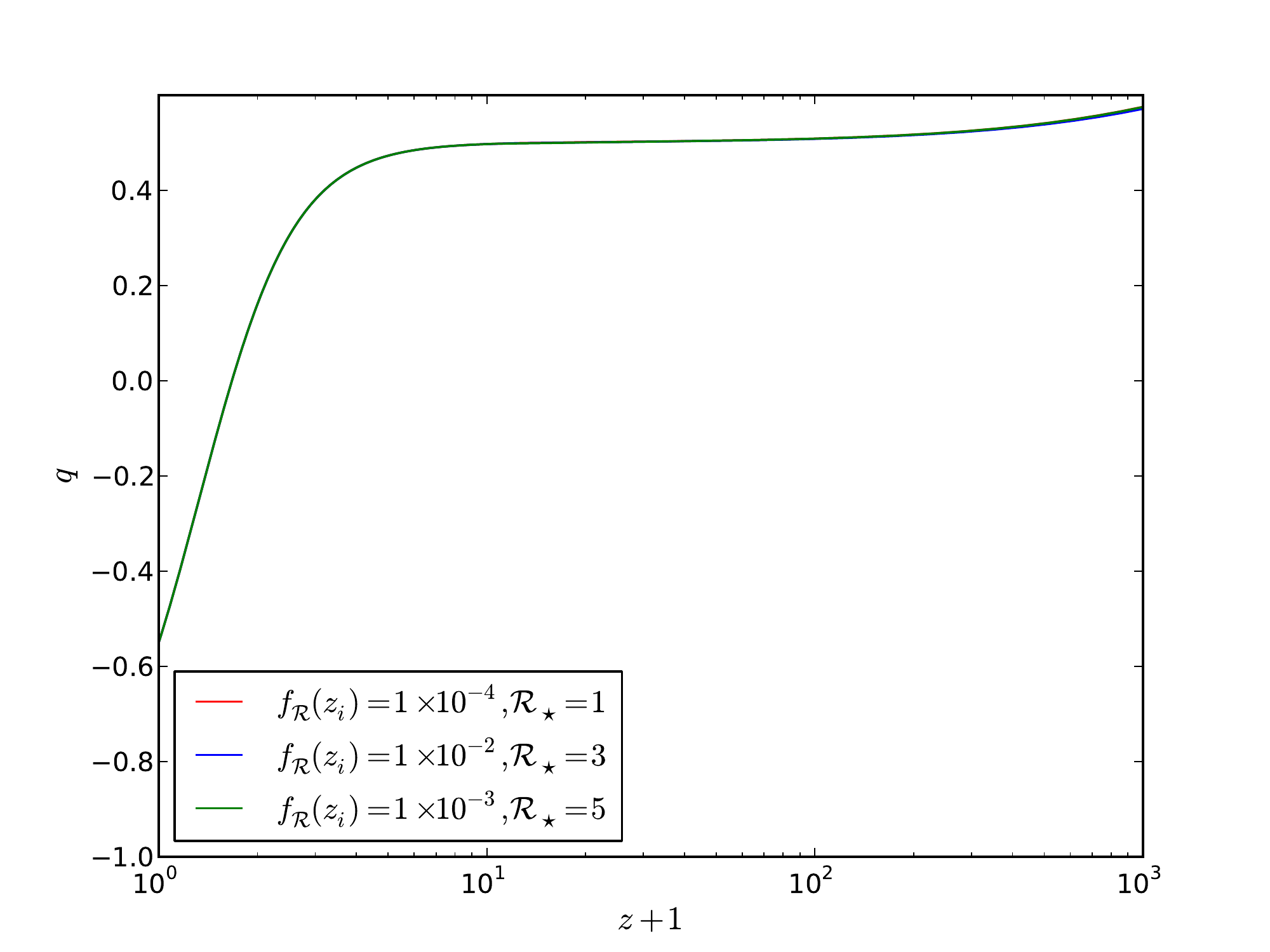} \\
\includegraphics[scale = 0.31]{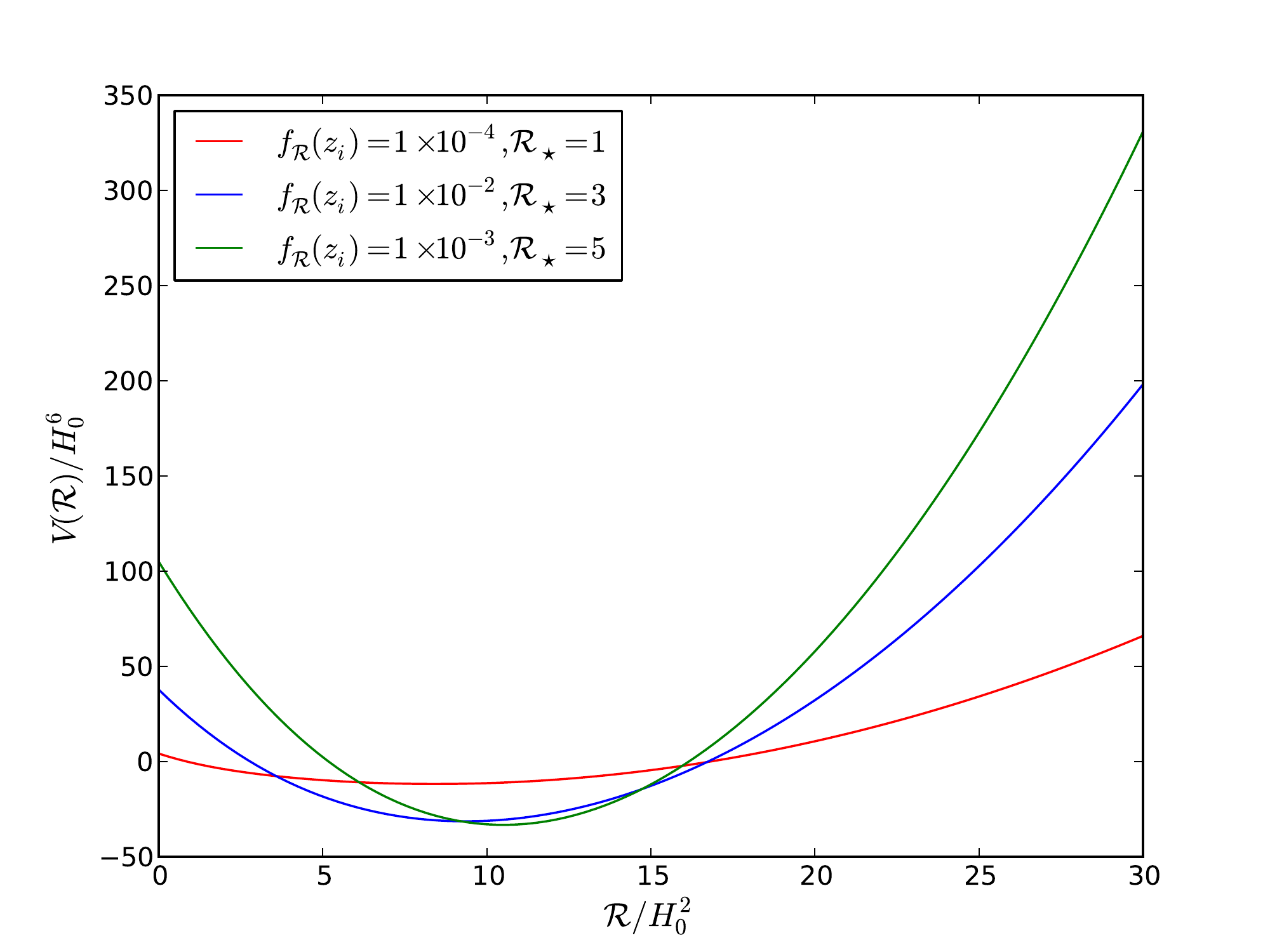}
\end{array}$
\end{center}
\caption{\label{expo1} The background evolution predicted by the exponential $f(\mathcal{R})$ model compared to $\Lambda$CDM. We choose to plot $\mathcal{R}$ far into the future ($z \rightarrow -1$), to explicitly show that our solution asymptotically tends to the minimum of the potential, $V(\mathcal{R})$, which we plot as well. We also plot the deceleration parameter $q$. The present-day matter energy density, $\Omega_{\rm{m}}$, is set to $0.30$. $\mathcal{R}_{\star}$ is in units of $H_{0}^{2}$.}
\end{figure}

In Fig.~\ref{expo1}, we plot the background evolution predicted by this model for a set of parameters as a function of redshift, $z$, using the prescription described at the end of Sec.~\ref{II}. We use a Brent algorithm (see \cite{brent_ref}) to find the correct $\Lambda_{\star}$ value that recovers a flat cosmology. Hence, the only true free parameters will be $\mathcal{R}_{\star}$ and $f_{\mathcal{R}}(z_{\rm{i}})$.

As we can see in Fig.~\ref{expo1}, the evolution of the Palatini Ricci scalar starts at a position where the potential is tilted and one would expect for it to roll down towards the minimum. However, at early times, the evolution of $\mathcal{R}$ is dominated by the stress--energy tensor trace or, equivalently, by the matter energy density. Since the ratio $f_{\mathcal{R}}/f_{\mathcal{R}\mathcal{R}}$ is negative throughout the whole evolution, the matter density contribution pushes $\mathcal{R}$ upwards the effective potential, while its slope and the Hubble friction term exert the opposite effect. As matter ceases to dominate close to the present, $\mathcal{R}$ inverts its motion and starts evolving towards the minimum, where it will asymptotically settle in the distant future.

Lastly, in Fig.~\ref{expo1}, we analyze the deceleration parameter $q$. We observe that our model predicts a Universe that will be expanding in an accelerated manner today, transitioning from a matter dominated decelerating phase at a redshift of around $z \approx 1$, as we have $q < 0$ at $z = 0$. This is a general result of this model, as can be inferred from the deceleration parameter equation 
\begin{equation}{\label{decelpar}}
 q = - \frac{\dot{H}}{H^{2}} - 1,
\end{equation}
where $\dot{H}$ is given by Eq.~\ref{hubbledotmain}. Switching from physical time $t$ to $\ln a$ we will have, at $z = 0$,
\begin{equation}{\label{hubblefrac}}
 q = -\frac{1}{6H_{0}^{2}} \PR{F(\mathcal{R})\mathcal{R} - 2 f(\mathcal{R}) + 3 H_{0}^{2} \Omega_{\rm{m}} - 12 H_{0}^{2}} - 1.
\end{equation}
As we observe in Fig.\ref{expo1}, our model evolves towards small values of $F(\mathcal{R})$ today, as the exponential is suppressed by the larger values of $\mathcal{R}$. Therefore, in Eq.~\ref{hubblefrac} we can neglect the $F(\mathcal{R}) \mathcal{R}$ term. Then, from Eq.~\ref{exponentialf}, we note that $f(\mathcal{R})$ will be dominated by $\Lambda_{\star}$. Since this parameter is determined from imposing a flat cosmology, $\Lambda_{\star} \approx - 2 \Lambda$, where $\Lambda \approx 2.1 H_{0}^{2}$ is the actual cosmological constant. In light of these arguments, Eq.~\ref{hubblefrac} becomes
\begin{equation}{\label{hubblefrac2}}
 q \approx - \frac{2\Lambda}{3 H_{0}^{2}} - \frac{\Omega_{\rm{m}}}{2} + 1 \approx -0.55.
\end{equation}
We verify in Fig.~\ref{expo1} that our prediction matches remarkably well the numerical values of obtained for $q$ today for different parameters of the exponential model. Hence, we conclude that this model should always predict an accelerated expansion today.

\subsection{\label{quad}The quadratic model}

The second model we introduce is the quadratic model, which we define by the function
\begin{equation}{\label{quadraticf}}
 f(\mathcal{R}) = \Lambda_{\star}\PC{1 + \mathcal{R}^{2}/\mathcal{R}_{\star}^{2}},
\end{equation}
\noindent where $\Lambda_{\star}$ and $\mathcal{R}_{\star}$ are a negative and positive constant of order $H_{0}^{2}$, just like in the previous model.

Computing the indefinite integral defined in Eq.~(\ref{effective_pot}) one can find the effective potential associated to this model:
\begin{equation}{\label{quadratic_pot}}
 V(\mathcal{R}) = -\frac{1}{3}\Lambda_{\star}\mathcal{R}^{2} - \frac{1}{9}\mathcal{R}^{3},
\end{equation}
\noindent which clearly has a global minimum at $\mathcal{R} = 0$ and a maximum at $\mathcal{R} = 2 \Lambda_{\star}$. Therefore, we expect the solution for $\mathcal{R}$ to asymptotically settle at the minimum in vacuum, leading to an effective cosmological constant value of
\begin{equation}{\label{effective_lamda_quad}}
 \Lambda_{\rm{eff}} \equiv \frac{1}{2} \PR{-2\Lambda_{\star}f_{\mathcal{R}}(0)-2f(0)} = -\Lambda_{\star}.
\end{equation}

\begin{figure}[t!]
\begin{center}$
\begin{array}{c}
\includegraphics[scale = 0.31]{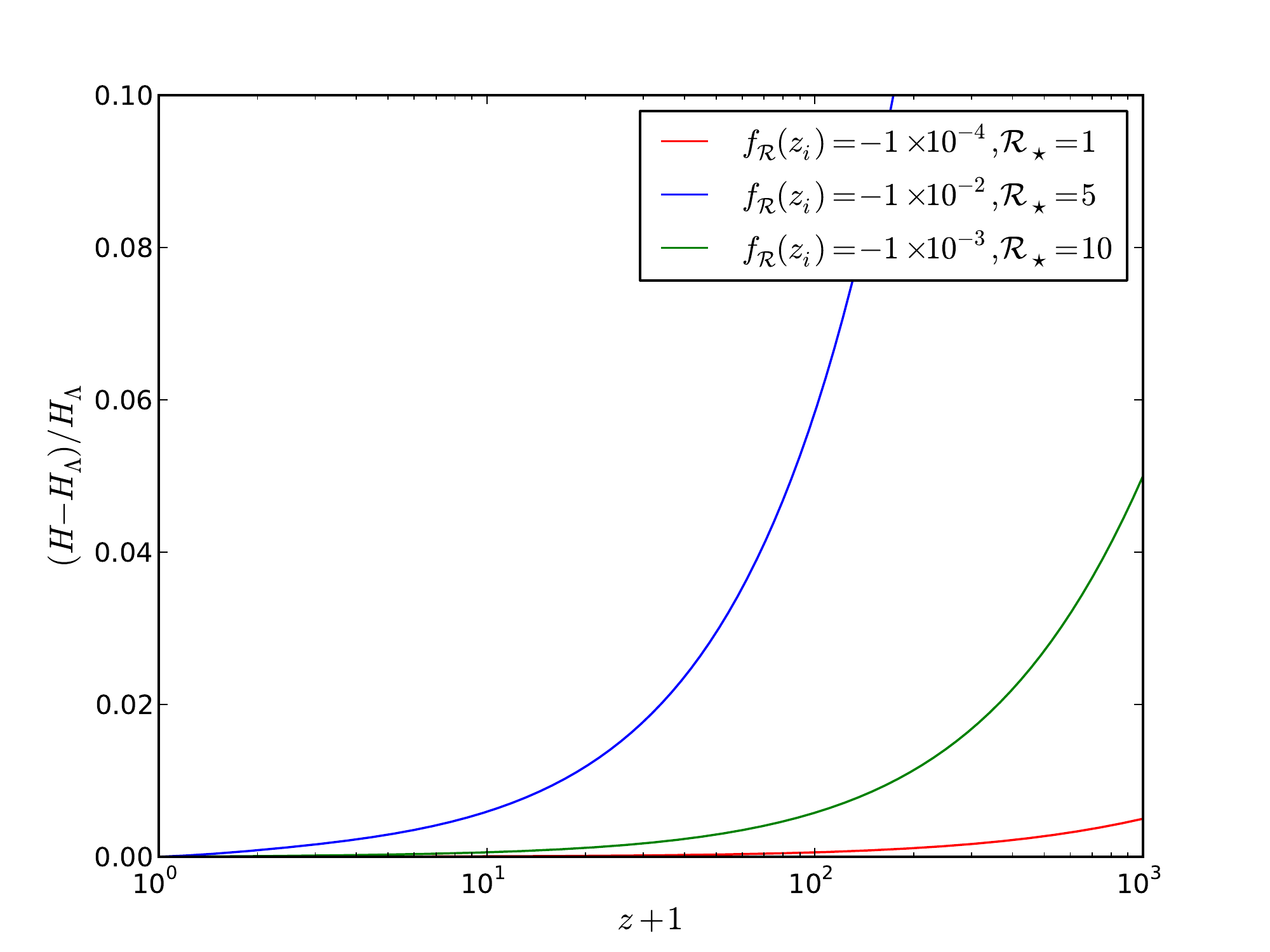} \\
\includegraphics[scale = 0.31]{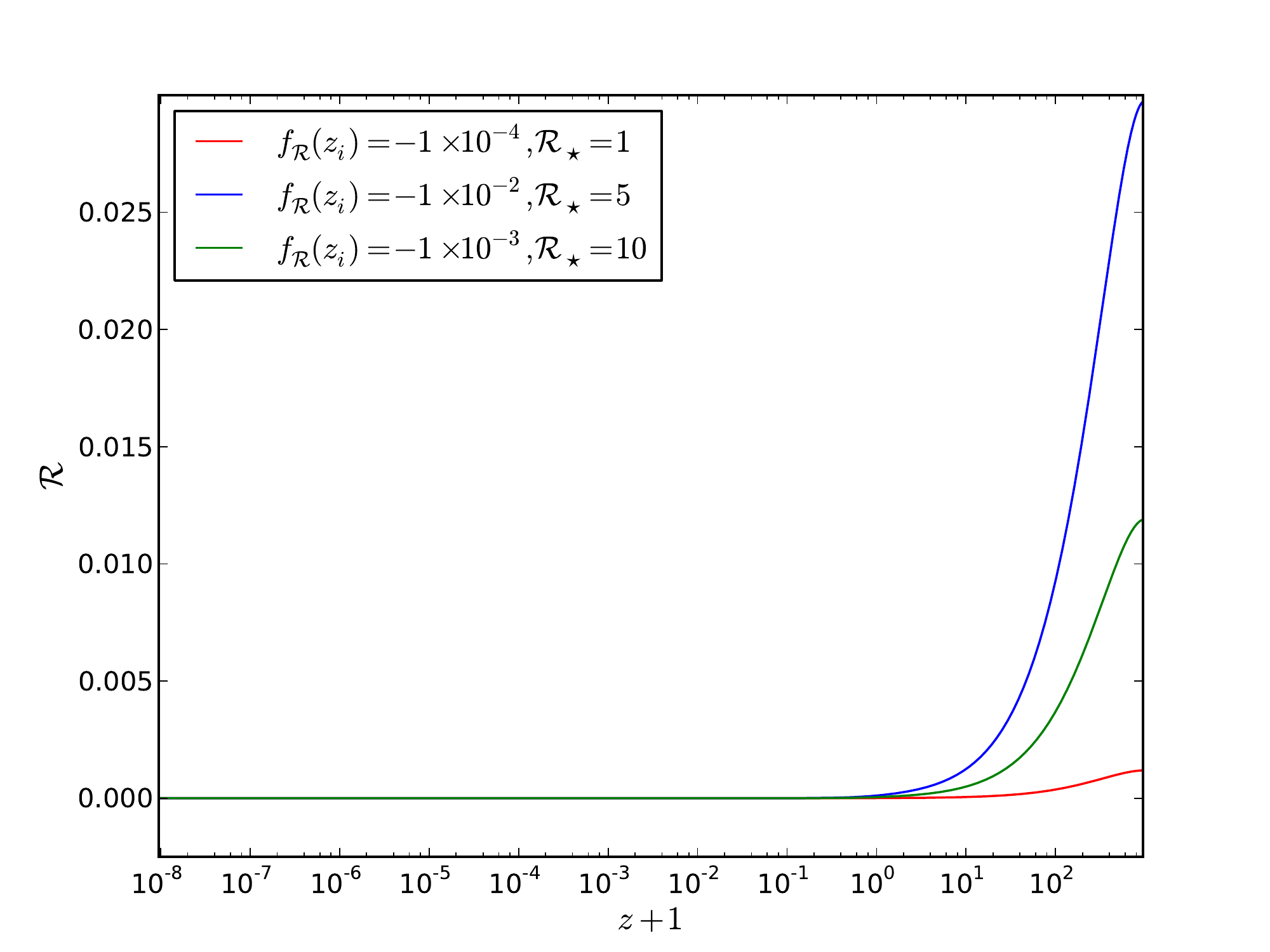} \\
\includegraphics[scale = 0.31]{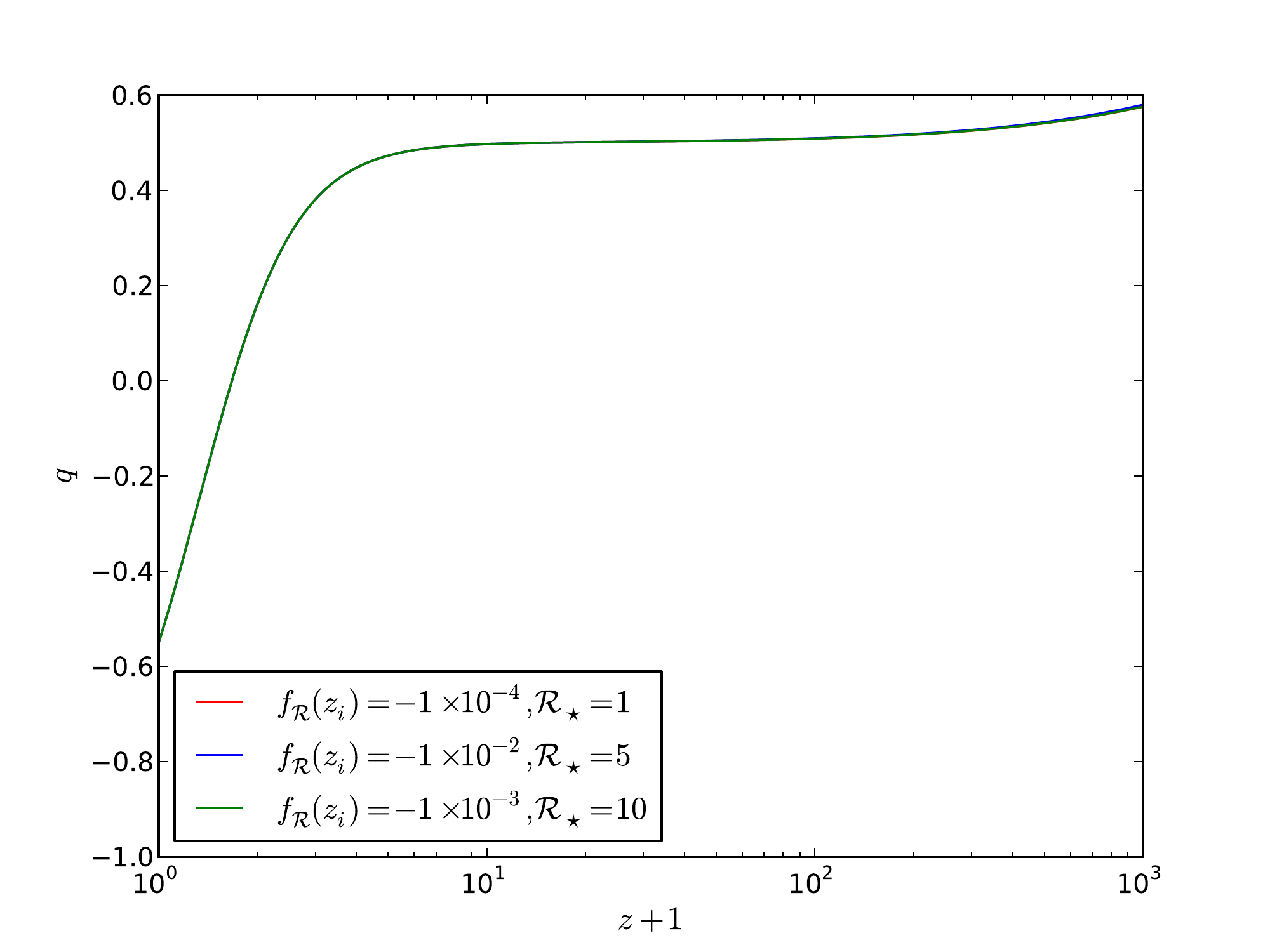} \\
\includegraphics[scale = 0.31]{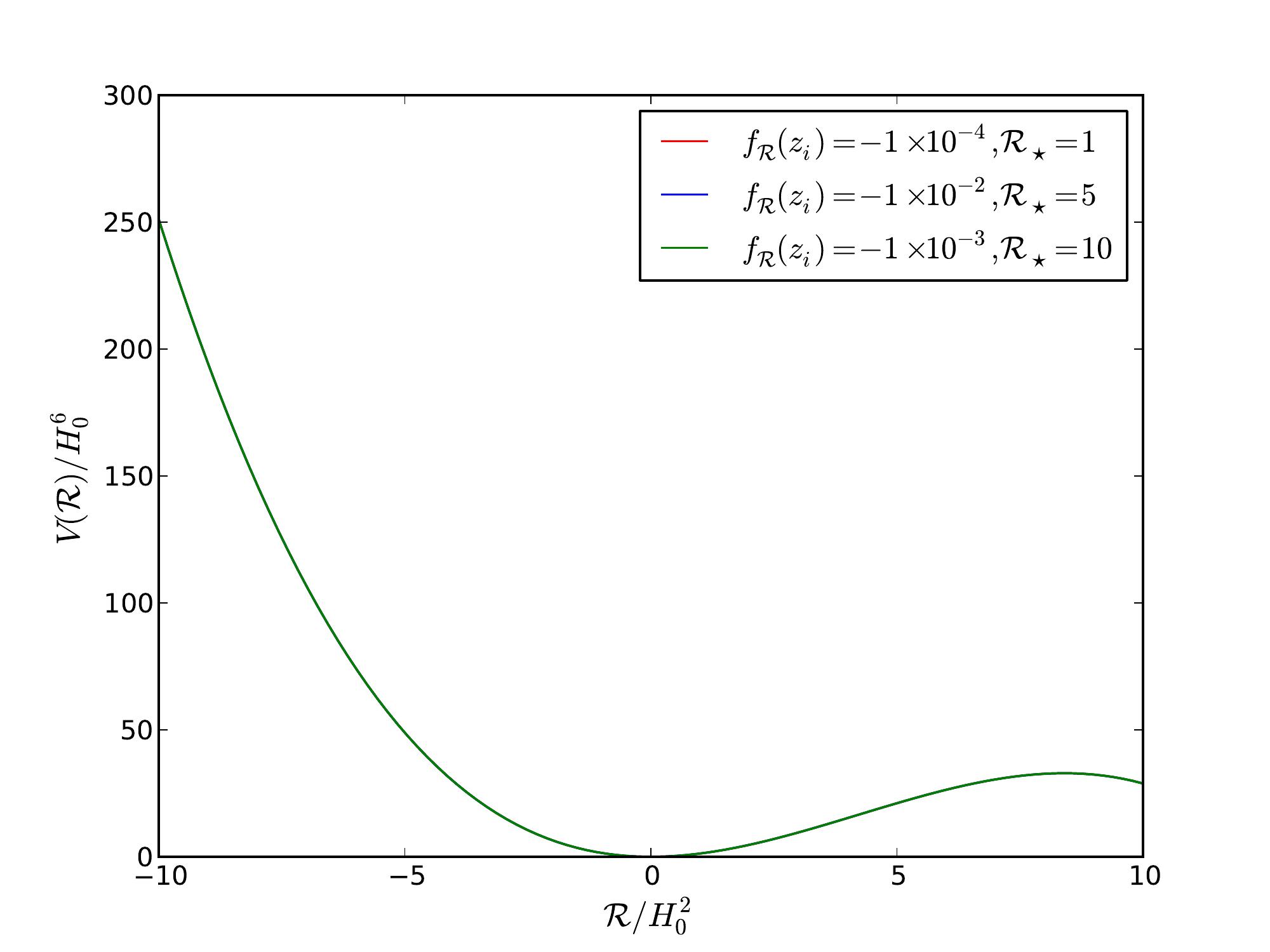}
\end{array}$
\end{center}
\caption{\label{quad1} The same as Fig.~\ref{expo1} for the quadratic $f(\mathcal{R})$ model. We have again fixed $\Omega_{\rm{m}} = 0.30$.}
\end{figure}

In Fig.~\ref{quad1}, we can see the evolution predicted by the quadratic model for $H$, compared to $\Lambda$CDM's, for different values of $f_{\mathcal{R}}(z_{\rm{i}})$ and $\mathcal{R}_{\star}$. We have chosen a negative $f_{\mathcal{R}}(z_{\rm{i}})$ as this sets the initial $\mathcal{R}$ to be positive.

In Fig.~\ref{quad1} we observe that $\mathcal{R}$ rolls down the effective potential towards the minimum from the beginning of the evolution, asymptotically settling there. When compared to the behavior seen for the exponential model in Fig.~\ref{expo1}, the different dynamics are linked to the ratio $f_{\mathcal{R}}/f_{\mathcal{R}\mathcal{R}}$ multiplying the stress--energy tensor in the dynamical equation for $\mathcal{R}$. For this particular setting of the quadratic model, this ratio is positive at early times, meaning that $\mathcal{R}$ will be pulled down by the matter energy density term towards the minimum of the potential, asymptotically settling there when the matter contribution becomes negligible. Had we chosen $\mathcal{R}(z_{\rm{i}})$ to be positive instead, the behavior wouldn't qualitatively change, with the matter term dictating $\mathcal{R}$ to evolve towards the minimum in a symmetric manner, starting from $\mathcal{R} < 0$.

Also, note that by setting the initial conditions for $f_{\mathcal{R}}(z_{\rm{i}})$ such that deviations from standard GR are kept small, $\mathcal{R}$ starts its evolution already close to the minimum, guaranteeing that it settles onto the minimum of the effective potential $V(\mathcal{R})$. Had we started the evolution such that the initial value of $\mathcal{R}$ would be beyond the maximum observed for the effective potential in Fig.~\ref{quad1}, $\mathcal{R}$ would roll indefinitely, jeopardizing the late-time achievement of an effective cosmological constant in this model.

Lastly, we have the evolution of the deceleration parameter $q$ in Fig.~\ref{quad1} predicted by this model. We observe that, close to the present, $f(\mathcal{R})$ will be dominated by $\Lambda_{\star}$ as can be seen in Eq.~\ref{quadraticf}, since $\mathcal{R}$ is tending to the minimum of the potential $V(\mathcal{R})$ located at zero. Hence, the analysis of the parameter $q$ in Sec.~\ref{expo} applies in this case as well. Therefore, this model should always predict an accelerating Universe today.

\section{BACKGROUND CONSTRAINTS}{\label{IV}}

\subsection{Observables}{\label{surveys}}

To constrain our models parameters, we perform a Metropolis-Hastings analysis using several background-related observables: the luminosity distance from Supernova type Ia (SNIa); the baryon acoustic oscillations (BAO); the shift parameter, acoustic scale and redshift of decoupling from the cosmic microwave background (CMB). For a detailed explanation on the statistics treatment of SNIa, BAO and CMB see \cite{Dodelson}.

\subsubsection{SNIa luminosity--redshift relation}

For the supernovae analysis, we have used the Union $2.1$ SNIa catalog from the ``Supernova Cosmology Project'' (SCP) \cite{supernova}. This data set is a compilation of $580$ type Ia Supernovae located over the redshift interval $0.623 < z < 1.415$. In our Metropolis-Hastings analysis, we have marginalized over the nuisance parameter $H_{0}$ using the procedure described in the appendix \cite{h0marg}. 

\subsubsection{Baryon Acoustic Oscillations peak}

Following the Planck analysis \cite{planckbao}, we have used BAO observations from the $6$dF Galaxy Redshift Survey ($6$dFGRS) at low redshift, $d^{\rm{obs}}(z = 0.106) = 0.336 \pm 0.015$ \cite{6rdf}, from the Luminous Red Galaxies (LRG) sample of the Sloand Digital Sky Survey (SDSS) $7$-year data release at the median redshift, $d^{\rm{obs}} (z = 0.35) = 0.1126 \pm 0.0022$ \cite{sdss7}, and the BAO signal from BOSS CMASS DR9, $d^{\rm{obs}} (z = 0.57) = 0.0732 \pm 0.0012$ \cite{boss}. 

\subsubsection{Cosmic Microwave Background}

\begin{table}[t!]
\caption{Inverse covariance matrix for the distance information obtained from Planck in the $\Lambda$CDM framework.}\label{table1}
\begin{center}
\begin{tabular}{c|c|ccc}
\hline\hline
\multicolumn{5}{c}{Planck}\\
\hline
&Best fit&$l_A(z_{\ast})$&$R(z_{\ast})$&$z_{\ast}$\\
\hline
$l_A(z_{\ast})$ & $301.77$ & $44.077$ & $-383.927$ & $-1.941$\\
$R(z_{\ast})$ & $1.7477$ & $-383.927$ & $48976.330$ & $-630.791$\\
$z_{\ast}$ & $1090.25$ & $-1.941$ & $-630.791$ & $12.592$\\
\hline\hline
\end{tabular}
\end{center}
\end{table}

In this work, we use the Planck distance information for the $\Lambda$CDM model to constrain our models. In table \ref{table1} it is possible to see the inverse covariance matrix and best-fit values obtained from Planck \cite{priors}.

\subsection{Metropolis - Hastings Algorithm}{\label{metro}}

Having described the procedure to calculate the $\chi^2$ of each observable we consider in the previous section, we continue towards the calculation of the confidence contours by means of a Metropolis--Hastings algorithm, which is a Markov Chain Monte Carlo method based on a stochastic sampling technique~\cite{2014arXiv1408.4438M}. One of the main advantages of the Metropolis--Hastings algorithm is its treatment of marginalized variables. When considering a subset of the parameters that form a chain, the marginalization over the remainder of the parameters occurs immediately, therefore making the treatment of the chains a simple process 


\subsection{Priors}{\label{priors}}

In this section we discuss the range of variation of the parameters we use in our models, which are the flat priors we have chosen to impose over them. Our two $f(\mathcal{R})$ models have a set of three parameters they share. These are the present-day relative energy densities of matter and baryons, $\Omega_{\rm{m}}$ and $\Omega_{\rm{b}}$, and the present-day value of the Hubble parameter, $H_{0}$. The ranges we have chosen for them are, respectively, $[0.01,0.99]$, $[0.001,0.080]$ and $[40.0,100.0]$.

Also for both models, $\Lambda_{\star}$ is determined by the background evolution, assuming we recover a flat cosmology today, at $a = 1$. As described in Sec.~\ref{expo}, this is achieved using a Brent algorithm (see \cite{brent_ref}) which ensures that the present-day value of the numerical Hubble parameter we obtain coincides with $H_{0}$. We start our background evolution at a redshift of $z_i \approx 1 \times 10^8$, and neglect any relativistic effects. Hence, we take the equation of state for matter to be zero throughout, i.e. $\rho_{\rm{m}} \propto a^{-3}$.

For the exponential model, we study two situations. In the first case, we choose to keep $f_{\mathcal{R}}$ at $z_{\rm{i}}$ fixed to a very small value, $10^{-4}$, and let $\mathcal{R}_{\star}$ vary between $[1.0,10.0]$. We do so to test the data against a definitive modification of gravity where the $\Lambda$CDM limit is not explicitly attainable.

In the second case, we let $f_{\mathcal{R}}(z_{\rm{i}})$ vary between $[1 \times 10^{-6}, 0.1]$ and $\mathcal{R}_{\star}$ between $[0.01,15.0]$. We can not consider $f_{\mathcal{R}}(z_{\rm{i}}) = 0$ due to the way we set the initial conditions in this model, as this would lead to a logarithmic divergence. Therefore, we can not set the deviations from standard GR plus $\Lambda$CDM at early times to zero, but only asymptotically minimize them by taking $f_{\mathcal{R}}(z_i) \rightarrow 0^{+}$. Nevertheless, we can test how large a deviation should be possible by considering an upper limit for $f_{\mathcal{R}}(z_{\rm{i}})$ of order $1 \times 10^{-1}$.

Lastly, for the quadratic model, we fix $\mathcal{R}_{\star}$ to a chosen value and let $f_{\mathcal{R}}$ at $z_i$ vary between $[-0.1,0.1]$. We chose to fix $\mathcal{R}_{\star}$ since the effective potential on which $\mathcal{R}$ evolves is independent of $\mathcal{R}_{\star}$ in the quadratic model, as seen in Eq.~(\ref{quadratic_pot}). 

\subsection{Results}{\label{results}}

In this section, we present the marginalized $2$-d contours for the posterior probabilities distributions of our parameters which were calculated using a Metropolis-Hastings algorithm, as described in Sec.~\ref{metro}. The plots shown here exhibit the combined constraints of the three background surveys we described in Sec.~\ref{surveys}. These plots were done using the plotting functions available in CosmoMC \cite{cosmomc}.
\begin{figure}[t!]
\begin{center}$
\begin{array}{c}
\includegraphics[scale = 0.375]{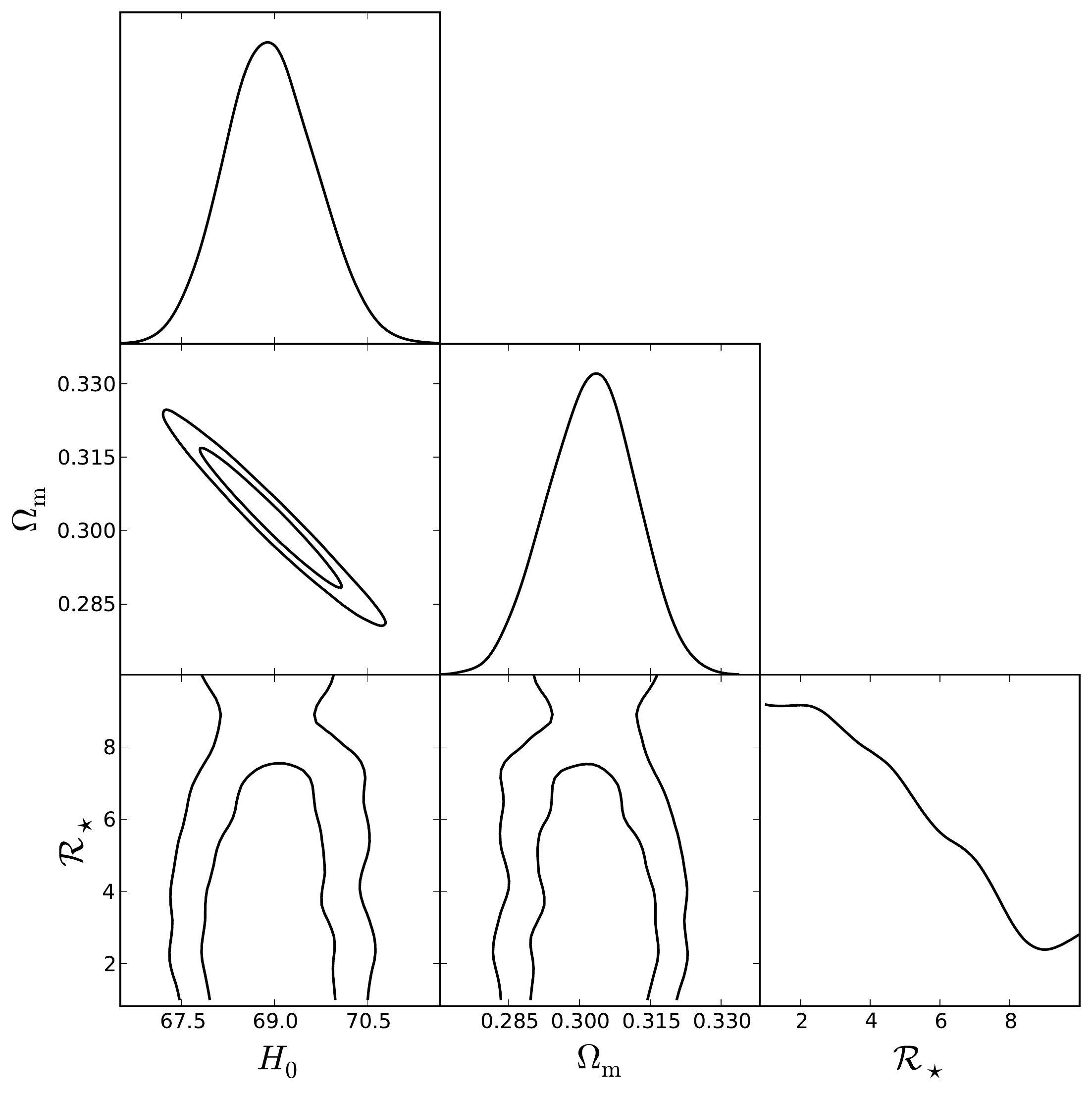}
\end{array}$
\end{center}
\caption{\label{contexpo} The $2$-d contours of the combined constraints from the background surveys we consider for the exponential $f(\mathcal{R})$ model with $f_{\mathcal{R}}(z_{\rm{i}})$ fixed to $10^{-4}$. We also plot the individual marginalized posterior probability distributions of each parameter.}
\end{figure}

\begin{figure}[t!]
\begin{center}$
\begin{array}{c}
\includegraphics[scale = 0.535]{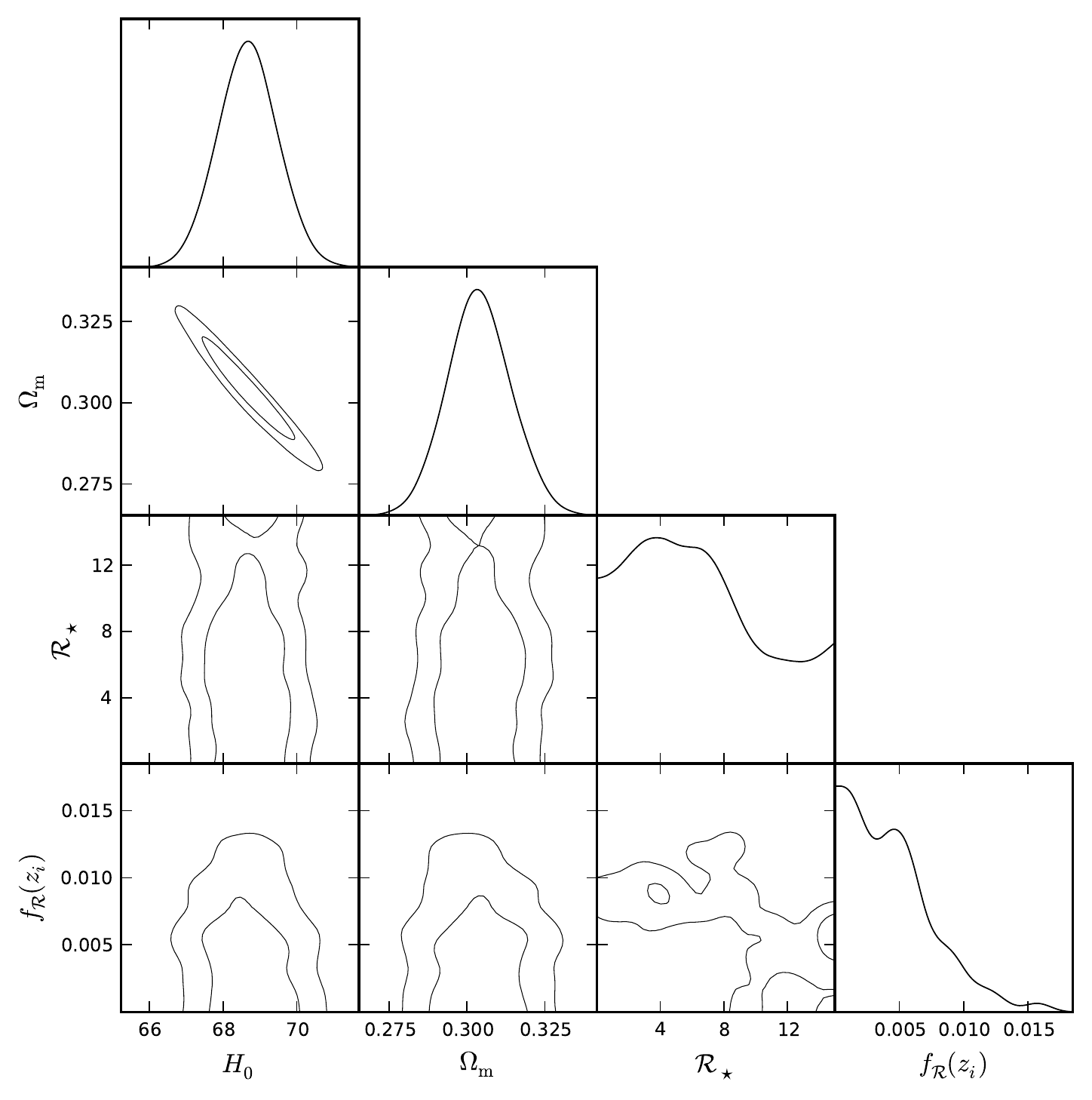}
\end{array}$
\end{center}
\caption{\label{contexponew} The $2$-d contours of the combined constraints from the background surveys we consider for the exponential $f(\mathcal{R})$ model by considering a prior range for $f_{\mathcal{R}}(z_{\rm{i}})$ between $[1\times10^{-6},0.1]$. We also plot the individual marginalized posterior probability distributions of each parameter.}
\end{figure}
In Fig.~\ref{contexpo} we have a triangular plot with the $2$-d contours between $H_{0}$, $\Omega_{\rm{m}}$ and $\mathcal{R}_{\star}$, as well with the individualized posterior distributions of each parameter for the exponential model considering $f_{\mathcal{R}}(z_{\rm{i}})$ fixed to $10^{-4}$. We obtain a slightly higher value of $H_{0}$ than the recent Planck result \cite{planckbao}, but we do not perform such a complete analysis, limiting ourselves to background observables. We also observe a smaller $\Omega_{\rm{m}}$ value than in \cite{planckbao}, which results from the combination of the different surveys we have considered, as the Union$2.1$ and BAO surveys do tend to prefer a slightly smaller $\Omega_{\rm{m}}$ value than Planck \cite{supernova,planckbao}. The $1$-$\sigma$ limits on these parameters are $H_{0} = 68.9 \pm 0.7$ and  $\Omega_{\rm{m}} = 0.303 \pm 0.009$.

For the $\mathcal{R}_{\star}$ parameter we cannot clearly state the confidence limits, as these are completely prior determined. Interestingly, we do observe a preference towards smaller values of $\mathcal{R}_{\star}$, possibly extending all the way to $0$ had we considered that limit. We set our initial conditions by imposing an initial $f_{\mathcal{R}}$ value at the starting redshift $z_{\rm{i}}$ that we invert to obtain the corresponding $\mathcal{R}_{\rm{i}}$. Hence, for this model, $\mathcal{R}_{\rm{i}} = - \mathcal{R}_{\star} \ln \PR{-f_{\mathcal{R}}(z_{\rm{i}}) \mathcal{R}_{\star}/\Lambda_{\star}}$. Therefore, we cannot have a pure GR plus $\Lambda$CDM for the exponential model because neither $f_{\mathcal{R}}(z_{\rm{i}})$ or $\mathcal{R}_{\star}$ can be set exactly to zero. However, the closest this model can get to $\Lambda$CDM, for a fixed $f_{\mathcal{R}}(z_{\rm{i}})$, is when $\mathcal{R}_{\star} \rightarrow 0^{+}$: in this limit, we observe that $\mathcal{R}_{\rm{i}}$ tends to decreasingly 
smaller values as $\mathcal{R}_{\star} \rightarrow 0^{+}$, while keeping the $\mathcal{R}/\mathcal{R}_{\star}$ ratio considerably large such that $f(\mathcal{R}) \rightarrow \Lambda_{\star}$. Hence, we recover an almost $\Lambda$CDM like evolution, which can be understood looking at the trace equation, Eq.~(\ref{einsteintrace}), which tends increasingly closer to the GR plus $\Lambda$CDM limit of $R + \kappa^2 T = 4 \Lambda$.
\begin{figure}[t!]
\begin{center}$
\begin{array}{c}
\includegraphics[scale = 0.375]{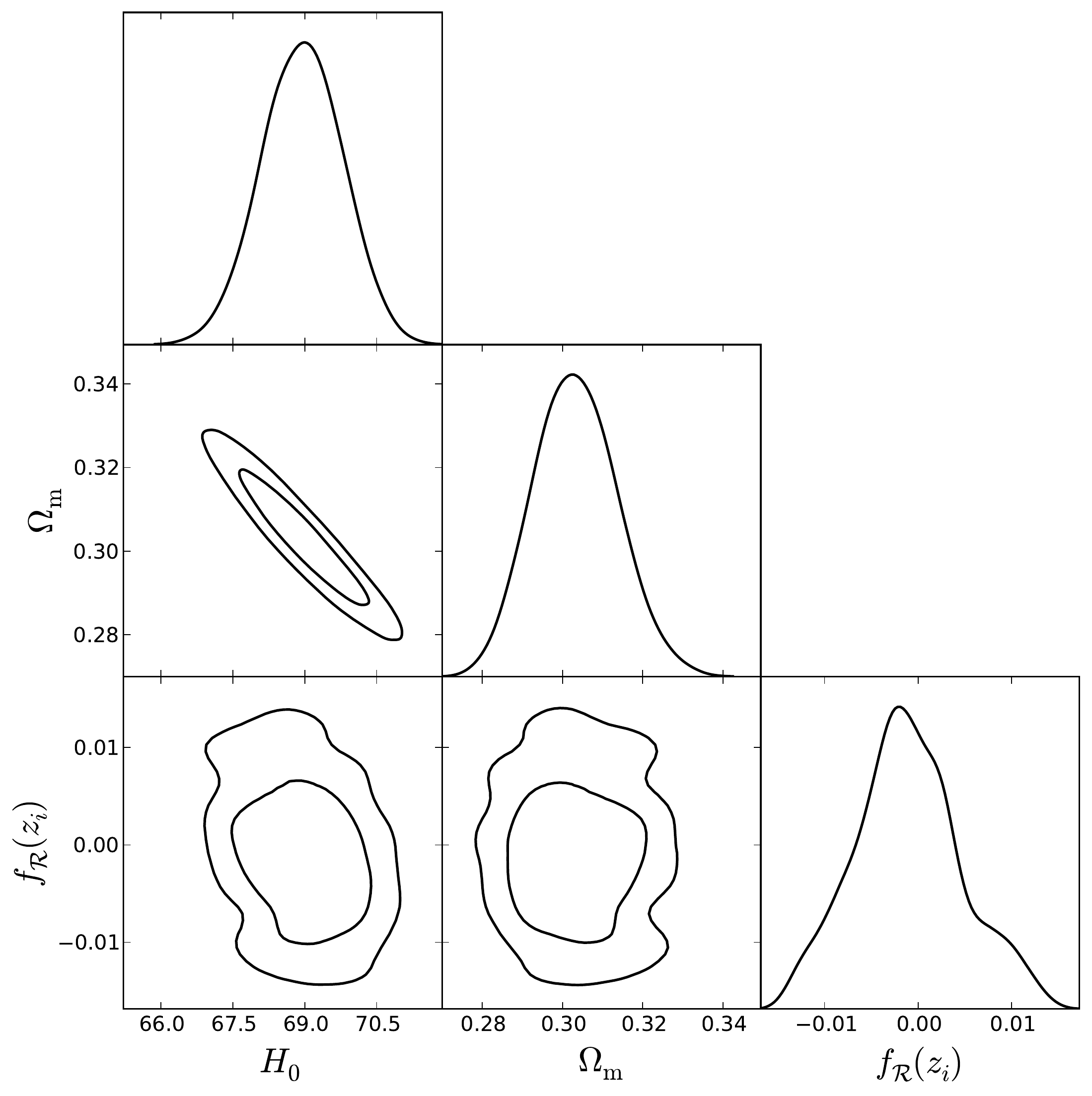}
\end{array}$
\end{center}
\caption{\label{contquad} Similar as Figs.~\ref{contexpo} and \ref{contexponew} for the quadratic $f(\mathcal{R})$ model.}
\end{figure}

Fig.~\ref{contexponew} presents the combined constraints on the exponential model considering the flat prior on $f_{\mathcal{R}}(z_{\rm{i}})$ between $[1 \times 10^{-6},0.1]$. We can observe a clear upper limit on $f_{\mathcal{R}}(z_{\rm{i}})$ of around $10^{-2}$, limiting, therefore, the maximum deviation from the actual gravitational constant $G$ we can have at early times. However, now $\mathcal{R}_{\star}$ appears even more unconstrained by the data, as larger values of $\mathcal{R}_{\star}$ are also allowed since these can be compensated by the $f_{\mathcal{R}}(z_{\rm{i}}) \rightarrow 0^{+}$ values: this limit pushes the model closer to the standard GR plus $\Lambda$CDM limit even for large values of $\mathcal{R}_{\star}$. We reinforce that we can not set the deviation from standard GR exactly to zero because that would lead to a logarithmic divergence in the initial conditions of this model. However, the lower limit we have chosen for the $f_{\mathcal{R}}(z_{\rm{i}})$ prior is much smaller than the 
current precision with which one can measure early deviations from the gravitational constant \cite{mgc1,mgc2}.

Lastly, in Fig.~\ref{contquad} we have the triangular plot with the confidence contours and $1$-d marginalized posterior probability distributions for the $f(\mathcal{R})$ quadratic model. We obtain very similar results for the standard cosmological parameters $H_{0}$ and $\Omega_{\rm{m}}$ as those observed in the exponential model in both situations, with the $1$-$\sigma$ limits on them being $H_{0} = 68.9 \pm 0.8$ and  $\Omega_{\rm{m}} = 0.303 \pm 0.010$.

For the quadratic model, we chose to keep $\mathcal{R}_{\star}$ fixed because it would not have a significant impact on the background evolution predicted by this model since it does not alter the shape of the potential on which $\mathcal{R}$ evolves. Hence, the third parameter we show constraints for is the initial value we impose for $f_{\mathcal{R}}$ at the starting redshift, $z_{\rm{i}}$. This, as detailed before, sets the maximum deviation from the standard gravitational constant $G$ one can have at early times in this model, since $f_{\mathcal{R}}$ evolves asymptotically to $0$ from its starting value.

It is clear that as $f_{\mathcal{R}}(z_{\rm{i}}) \rightarrow 0$ we get closer to a $\Lambda$CDM like evolution. Given it would be numerically hard to evolve the model if $f_{\mathcal{R}}(z_{\rm{i}}) = 0$, we made the approximation that, if the Metropolis--Hastings algorithm encountered such a value, we would have an exact $\Lambda$CDM background evolution. The results we have obtained show a preference for a standard GR plus $\Lambda$CDM scenario, as can be seen in the $1$-d posterior probability distribution for $f_{\mathcal{R}}(z_{\rm{i}})$ in Fig.~\ref{contquad}. The corresponding $1$-$\sigma$ confidence limits are $f_{\mathcal{R}}(z_{\rm{i}}) = -0.001 \pm 0.006$. We also observe a symmetry on the posterior distribution of this parameter, which could be expected given that the evolution of $\mathcal{R}$ is symmetric under the change of sign of $f_{\mathcal{R}}$ for, as discussed in Sec.~\ref{quad}. 

\section{CONCLUSION}{\label{conclusion}}

In this work, we explored a way to obtain the background evolution for two different models of the novel hybrid metric-Palatini theory of gravity. We re-wrote the dynamical equation for the additional degree of freedom introduced by this theory as a dynamical equation for the actual Palatini Ricci scalar $\mathcal{R}$. We define the initial conditions by imposing the deviation one has from standard GR at early times. Hence, we set a small value for $f_{\mathcal{R}}(z_{\rm{i}})$ and invert the latter to obtain $\mathcal{R}(z_{\rm{i}})$, while keeping $\dot{\mathcal{R}}(z_{\rm{i}}) = 0$.

We define an effective potential $V(\mathcal{R})$ where the Palatini Ricci scalar evolves and, if a minimum exists, $\mathcal{R}$ should asymptotically settle there in vacuum, so that one recovers standard GR plus an effective cosmological constant at late times. $V(\mathcal{R})$ could potentially have a complicated form. However, for the models we introduce here, that is not the case.

We present the exponential and quadratic $f(\mathcal{R})$ models and show that the background evolution predicted by them does not deviate much from $\Lambda$CDM. This could be different, of course, had we decided to set the deviation from the gravitational constant $G$ in the high-redshift regime to be large. Also, we explicitly show the effective potential $V(\mathcal{R})$ for both models and $\mathcal{R}$ asymptotically tending to its minimum at late times. This is less obvious in the exponential model as the matter term in Eq.~(\ref{dynamicalpala}) initially drives the Palatini Ricci scalar up the potential, only for it to later slowly fall down towards the minimum due to the potential slope.

We also study and present the evolution of the deceleration parameter $q$ for our models. We verify, numerically, that they predict an accelerated expansion today for different values of their parameters. Furthermore, we perform a simple analytical analysis of $q$ and conclude that our models yield a present-day value for it around $-1/2$, meaning they are suitable candidates for producing cosmological acceleration today.

We then use background CMB, BAO and Supernovae data to constrain the models. Keeping $f_{\mathcal{R}}(z_{\rm{i}})$ fixed to $10^{-4}$ for the exponential model,  we cannot state an actual constraint on the $\mathcal{R}_{\star}$ parameter. However, we do note how the data, as expected, seems to tend towards the $\Lambda$CDM limit. We believe that, had we not chosen to restrict the prior range in order to have a definitive modification of gravity without the $\Lambda$CDM limit, we would see the lower range of our confidence contours in Fig.~\ref{contexpo} tending to $0^{+}$ in $\mathcal{R}_{\star}$.

Still for the exponential model, when we impose a flat prior on $f_{\mathcal{R}}(z_{\rm{i}})$ between $[1\times10^{-6},0.1]$ we observe a clear upper limit of order $10^{-2}$. This value marks the maximum deviation one can have at early times from the actual gravitational constant $G$. Also, the data exhibits a marked tendency towards the lower limit of $f_{\mathcal{R}}(z_{\rm{i}})$, as $f_{\mathcal{R}}(z_{\rm{i}}) \rightarrow 0^{+}$ minimizes the early time deviations from the gravitational constant $G$, allowing the model to get asymptotically closer to the standard GR+$\Lambda$CDM limit.

For the quadratic model, we constrained the initial value of $f_{\mathcal{R}}$ while keeping $\mathcal{R}_{\star}$ fixed which means that, effectively, we are again constraining the maximum deviation one has from the gravitational constant at early times. As expected, the results indicate a preference towards no deviation at all, as in the standard GR plus $\Lambda$ limit. We obtain $f_{\mathcal{R}}(z_{\rm{i}}) = -0.001 \pm 0.006$ as the confidence limits for our parameter.

Hence, we see in this work that $f_{\mathcal{R}}(z_{\rm{i}})$ could play an important role in the constraining of this theory, as it sets the deviation one observes from standard GR at early times. The background data we have used permits a deviation around $1\%$ from the actual gravitational constant $G$ at early times, which is within constraints on $G_{\rm{eff}}/G$ coming from BBN and CMB data \cite{mgc1,mgc2}.

This, combined with the fact that the Newtonian potentials also exhibit a departing behavior from $\Lambda$CDM at early times \cite{hybridpert}, suggests that it would be very interesting to constrain these models using the latest Planck data available by enhancing the formalism developed so far with the inclusion of perturbation observables.

\acknowledgments

N.A.L.\ acknowledges financial support from Funda\c{c}\~{a}o para a Ci\^{e}ncia e a Tecnologia (FCT) through grant SFRH/BD/85164/2012. V.S.-B.\ acknowledges funding provided by CONACyT and the University of Edinburgh. We thank Andrew Liddle for useful comments and discussions.

\bibliography{hybridback}

\end{document}